\documentclass[conference]{IEEEtran}
\IEEEoverridecommandlockouts

\usepackage{textcomp}
\usepackage[T1]{fontenc}

\usepackage{amsmath,amstext,amsfonts}
\usepackage{amssymb}
\usepackage{mathtools}
\usepackage[cmintegrals]{newtxmath}
\usepackage{bm}  
\usepackage{siunitx}
\usepackage{physics}
\DeclareMathOperator*{\argmin}{arg\,min}

\usepackage{graphicx}
\graphicspath{%
	{Figures/}
	{Figures/TiKz/}
	{Figures/Ipe/}
	{Figures/XFig/}
	{Figures/Inkscape/}
	{Figures/Matlab/}
	{Figures/Scanned/}
	{Figures/Saved/}
}

\usepackage[list=true,labelformat=simple]{subcaption}
\usepackage{cite}
\usepackage{url}
\usepackage{xcolor}

\usepackage{booktabs}  

\usepackage[inline]{enumitem}

\usepackage{algorithm,algpseudocode,float}
\usepackage{setspace} 

\algnewcommand{\algorithmicand}{\textbf{ and }}
\algnewcommand{\algorithmicor}{\textbf{ or }}
\algnewcommand{\AlgAnd}{\algorithmicand}
\algnewcommand{\AlgOr}{\algorithmicor}

\newsavebox{\myparbox}
\newlength{\myparboxwidth}

\newcommand{\degree}{\ensuremath{^\circ}} 
\providecommand{\abs}[1]{\lvert#1\rvert}
\providecommand{\norm}[1]{\lVert#1\rVert}

\newcommand\Angle[1]{\setbox0=\hbox{$\mskip 7mu minus 4mu#1$}%
  \raise.21ex\hbox{$/$}\hskip-0.95ex\underline{\raise\dp0\hbox{\box0}}}

\ExplSyntaxOn

\NewDocumentCommand \vect { s o m }
 {
  \IfBooleanTF {#1}
   { \vectaux*{#3} }
   { \IfValueTF {#2} { \vectaux[#2]{#3} } { \vectaux{#3} } }
 }

\DeclarePairedDelimiterX \vectaux [1] {\lbrack} {\rbrack}
 { \, \dbacc_vect:n { #1 } \, }

\cs_new_protected:Npn \dbacc_vect:n #1
 {
  \seq_set_split:Nnn \l_tmpa_seq { , } { #1 }
  \seq_use:Nn \l_tmpa_seq { \enspace }
 }
\ExplSyntaxOff

\usepackage[firstpageonly=true]{draftwatermark}
\usepackage{hyperref}
\hypersetup{
    colorlinks=true,
    linkcolor=blue,
    filecolor=magenta,
    urlcolor=cyan,
    bookmarks=true,
}
\usepackage[noabbrev]{cleveref}  
\Crefname{figure}{Fig.}{Figs.}

\def\BibTeX{{\rm B\kern-.05em{\sc i\kern-.025em b}\kern-.08em
    T\kern-.1667em\lower.7ex\hbox{E}\kern-.125emX}}
\begin{document}

\title{A Policy Iteration Approach for Flock Motion Control
\thanks{This work was partially supported by NSERC Grant~EGP~537568-2018.}
}

%

%
\author{%
  \IEEEauthorblockN{Shuzheng Qu\IEEEauthorrefmark{1}, Mohammed Abouheaf\IEEEauthorrefmark{2}, Wail Gueaieb\IEEEauthorrefmark{1} and Davide Spinello\IEEEauthorrefmark{3}}
  \IEEEauthorblockA{\IEEEauthorrefmark{1}School of Electrical Engineering and Computer Science\\
    \IEEEauthorrefmark{3}Department of Mechanical Engineering\\
    University of Ottawa, Ottawa, Ontario, Canada K1N~6N5\\
    Email: \{fqu096, wgueaieb, dspinell\}@uottawa.ca}
  \IEEEauthorblockA{\IEEEauthorrefmark{2}College of Technology, Architecture \& Applied Engineering\\
    Bowling Green State University, Bowling Green, 43402, OH, USA\\
    Email: mabouhe@bgsu.edu}
}

\maketitle

\begin{abstract}
The flocking motion control is concerned with managing the possible conflicts between local and team objectives of multi-agent systems. The overall control process guides the agents while monitoring the flock-cohesiveness and localization. 
The underlying mechanisms may degrade due to overlooking the unmodeled uncertainties associated with the flock dynamics and formation.
On another side, the efficiencies of the various control designs rely on how quickly they can adapt to different dynamic situations in real-time. An online model-free policy iteration mechanism is developed here to guide a flock of agents to follow an independent command generator over a time-varying graph topology. 
The strength of connectivity between any two agents or the graph edge weight is decided using a position adjacency dependent function. 
An online recursive least squares approach is adopted to tune the guidance strategies without knowing the dynamics of the agents or those of the command generator. It is compared with another reinforcement learning approach from the literature which is based on a value iteration technique. The simulation results of the policy iteration mechanism revealed fast learning and convergence behaviors with less computational effort. 
\end{abstract}


\DraftwatermarkOptions{%
angle=0,
hpos=0.5\paperwidth,
vpos=0.97\paperheight,
fontsize=0.012\paperwidth,
color={[gray]{0.2}},
text={
  \newcommand{\thispaperdoi}{10.1109/ROSE52750.2021.9611776}
  \newcommand{\thispaperCopyrightYear}{2021}
  \parbox{0.99\textwidth}{This is the postscript version of the published paper. (doi: \href{http://dx.doi.org/\thispaperdoi}{\thispaperdoi})\\
    \copyright~\thispaperCopyrightYear~IEEE.  Personal use of this material is permitted.  Permission from IEEE must be obtained for all other uses, in any current or future media, including reprinting/republishing this material for advertising or promotional purposes, creating new collective works, for resale or redistribution to servers or lists, or reuse of any copyrighted component of this work in other works.}},
}

\section{Introduction}

Cooperative control schemes can sometimes be challenged by the increasing complexity of the dynamics and connectivity of their underlying systems~\cite{Grundel2007}.
This can be manifested in many applications where conflicting objectives may coexist, such as fleets of self-driving vehicles, unmanned aircraft, and large warehouse robots~\cite{Zhihua2008, Wurman2008}.
The autonomy  of such interacting systems relies on how well the pre-tuned control strategies can cope with the unmodeled dynamics or even perform under unexpected operational scenarios. 
Many distributed control approaches have been investigated and widely adopted in this domain~\cite{ICRA1,ICRA2}. The flocking motion control problem is one category of these problems where each agent decides its action based on a compromise between a set of competing, and possibly conflicting, objectives~\cite{Speck2018}. In the context of the current manuscript, The objectives of the flock control scheme can be summarized as the guidance of agents towards a common goal, preventing collision among the flock by imposing a minimum safety distance between the agents, and the cohesion of the flock members towards a common speed~\cite{gu2008using,abouheaf2017flocking}. Herein, a model-free Reinforcement Learning (RL) process is used to develop a control strategy to guide agents communicating over a time-varying graph topology. Further, the control strategies are adapted online in real-time without prior knowledge of the agent dynamics. This research integrates ideas from optimal control, adaptive control, graph theory, reinforcement learning, and fuzzy logic to solve the aforementioned flocking motion control problem.

The study of multi-agent systems pertains to the behavior analysis of a large number of agents, such as flocks of flying birds and fish schools. Since the early works in~\cite{reynolds1987flocks,herd}, researchers have been reflecting back these ideas to develop distributed control approaches for cooperative systems. 
Distributed estimators have been employed to solve a leader-follower tracking problem under noisy environments in~\cite{hu2010distributed}. Sliding surface control is a common approach that is used to handle the coordination tasks within multi-agent systems~\cite{hu2010distributed}. 
Adaptive control mechanism is used to trace the peaks of unknown fields for a multi-agent system under uncertain environments in~\cite{jadaliha2010adaptive}.
A server-based approach is employed to control a group of embedded control systems in~\cite{aerial}. 
Another distributed control scheme is developed by combining consensus algorithms and attraction/repulsion potential functions in~\cite{fish}.
The consensus control problems for multi-agent systems communicating over fixed graph topologies are solved using pinning gain control ideas in~\cite{AbouheafAuto14,AbouheafCDC13,AbouheafIJCNN13}.

Machine learning provide useful tools like reinforcement learning, fuzzy systems, and neural networks which are employed to solve many adaptive control problems~\cite{sut92,AbouheafIRL2019}. 
The RL processes enable the agent to learn the best strategy-to-follow through interactions with its dynamic environment~\cite{Sutton_1998}. This is often done using one of two two-step techniques known as Value Iteration~(VI) and Policy Iteration~(PI). The implementation of these techniques is often performed using means of adaptive critics~\cite{Bertsekas1996}. PI processes are developed for systems interacting over graphs using actor-critic neural structures in~\cite{AbouheafCTT2015,Abouheapolicy2017}.  
On a relevant side, fuzzy logic is applied to solve control problems for systems of imprecise, vague, or uncertain dynamics. It provides not only a way of involving the designer experience in solving such problems but it can also lead to a balanced strategy for the conflict-resolution of multiple optimization criteria~\cite{Singh2013}. Fuzzy logic is employed to develop an obstacle avoidance scheme to solve a leader-follower problem  in~\cite{innocenti2007multi,6762934}. A fuzzy logic scheme based on an extended Takagi-Sugeno-Kang (TSK) inference system is introduced in a collision-avoidance task in a flock motion control application~\cite{abouheaf2017flocking}. A combined Fuzzy-RL system is tested in~\cite{ICRA21} for the online tuning of the rule consequences of a zero-order TSK fuzzy system.

The implementation of PI solutions using regression models, like the online and offline least squares methods, may result in unstable computational processes~\cite{Busoniu2010, Srivastava2019}. This problem becomes more challenging when a multi-agent system is considered where it is required to solve a set of coupled Bellman optimality equations simultaneously~\cite{Lewis2012}. Recursive least squares (RLS) regression approaches provide adequate mechanisms to handle such concerns taking into account the online processing of the sensory data feedback~\cite{Yin2015}. 
The RLS approach has been utilized in many applications, like in real-time signal processing and channel equalization~\cite{1315946, 1163587}, adaptive control~\cite{1186866}, and wireless communication~\cite{6529766}, to name a few. A generalized data estimator that is based on RLS is employed for data analysis, mining and prediction in~\cite{5451430}. The RLS technique exhibited better convergence features and faster parameter tracking abilities when compared with other gradient-based search techniques~\cite{1163587}. 
A distributed estimator based on a communication diffusion algorithm is developed to optimize a global RLS criteria for a multi-agent system in~\cite{8641452}. 
RLS is applied to tune neuro-fuzzy structures for different optimization applications and adaptive systems~\cite{YUSOF20111717, 7801936, 6083937, hu2005adaptive}. A self-tuning scheme based on RLS learning approach is used to adapt the fuzzy rules of a TSK fuzzy system in~\cite{7801936}. A generalized RLS approach is adopted to train a neural network structure in~\cite{1593689}.

The paper tackles the problem of flock motion control.
The contributions of this work are two-fold:
First, a PI approach is advised to solve a set of coupled Bellman optimality equations in real-time. This is accomplished to adapt tracking strategies in a distributed fashion without any prior knowledge of the agents' dynamics. Second, an algorithm based on RLS is employed to implement the PI solution. The proposed solution offers faster convergence features when compared with a similar method based on a VI process~\cite{ICRA21}. The current work considers a time-dependent connectivity graph among the agents, unlike the fixed connectivity assumed in~\cite{ICRA21}.
The remaining of the paper is organized as follows: \Cref{problem_fromulation} introduces the task in hand casted as an optimization problem along with a high-level overview of the proposed motion control approach. The control structure is described in details in \Cref{track,separation,cons}. \Cref{Sim} validates the proposed method through a number of test cases. Finally, \Cref{conc} offers some concluding remarks.

\section{Problem Formulation}
\label{problem_fromulation}
The problem in hand is to control the motion of $M$-agents communicating over an undirected time-varying graph topology. The objective is to guide the flock members to follow a leader while satisfying the following requirements simultaneously 
\begin{enumerate*}[before=\unskip{: }, itemjoin={{; }}, itemjoin*={{; and }}]
\item avoid collision between neighboring agents
\item adjust the velocities of agents to reach a consensus on common flock velocity.
\end{enumerate*}
The leader $\tau \in \{1,2,\dots, M\}$ could be an independent command source as well as a virtual dynamic trajectory.  The leader's dynamics are independent of those of the followers.
Hence, this problem is generalizable to different classes that range from guidance of unmanned vehicles to solving pursuer-evader games. 

The dynamics of each agent are approximated by
\begin{align*}
  \bm{g}_{k+1}^i & = \bm{g}_{k}^i + T \, {\bm{h}}_{k}^i
  &
    {\bm{h}}_{k+1}^i & = {\bm{h}}_{k}^i + T\, \bm{c}_k^i
\end{align*}
where $\bm{g}_{k}^i = [x^{i}_{k} \ y^{i}_{k}]^T \in \mathbb{R}^2$ and $\bm{h}_{k}^i = [v^{i,x}_{k} \ v^{i,y}_{k}]^T \in \mathbb{R}^2$ represent the position and linear velocity at discrete time step $k\ \in \ \mathbb{N}$, respectively, $T$ is the sampling period, and $\bm{c}_k^i = [c^{i,x}_{k}, c^{i,y}_{k}]^T \in \mathbb{R}^2$ is the control signal vector of agent~$i \in \{1,2,\dots, M\}$.
Agent \textit{i} receives different  measurements at time-step $k$ and hence generates a corresponding overall control signal $\bm{c}_k^i$.

The local objective of the flock members is to keep a desired safety distance between each agent~$i$ and its neighbors. The team objectives involve reaching a consensus on flock-velocity and that the agents maintains an average proximity $d_t$ from the leader.
The objectives can be formally described by the following relations for each agent~$i \in \{1,\ldots,M\} \setminus \{\tau\}$ and $\gamma \in \{x,y\}$:
\begin{subequations}
  \label{eq:objectives}
  \begin{align}
    \label{eq:objectives:track-leader}
    & \lim_{k \to \infty}  \qty\Big( \sum_{i \in \{1,\ldots,M\} \setminus \{\tau\}} \norm{ \bm{g}_{k}^i - \bm{g}_{k}^{\tau} } )  \Big/  \qty\Big(M-1) = d_t
    \\
    \label{eq:objectives:separation}
    & \lim_{k \to \infty} \abs{\gamma_k^i-\gamma_k^j} \geq s, ~\forall j \in {\cal M}_i
    \\
    \label{eq:objectives:velocity-consensus}
    & \lim_{k \to \infty} v_k^{i,\gamma} = v^{\sim,\gamma}
  \end{align}
\end{subequations}
where ${\cal M}_i$ is the set of neighborhood agents of agent~$i$, $s$ is the desired safety distance between each agent and its neighbors, and $v^{\sim,\gamma}$ is the flock's consensus speed. The significance of $\gamma$ is to articulate that the $x$ and $y$-components of the different signals are treated in a similar fashion. 
To reflect the system objectives, the control command of each agent~$i \in \{1,\ldots,M\} \setminus \{\tau\}$ is formulated as an aggregate of three auxiliary control signals as
\begin{gather}
  {c}^{i,\gamma}_{k}
  = {c}_{t,k}^{i,\gamma}
  + {c}_{v,k}^{i,\gamma}
  + {c}_{d,k}^{i,\gamma}
  .
  \label{conttotal}
\end{gather}
The tracking control signal $c_{t,k}^{i,\gamma}$ is decided as a function of the positions of each follower and leader, $ c_{t,k}^{i,\gamma} \ =\ f^i_{t,\gamma}( \gamma^{i}, \gamma^{\tau})$.
The signal $c_{v,k}^{i,\gamma}$ achieves a consensus on a common flock velocity in real-time. This signal is determined in terms of the positions and velocities of each agent and those of its neighbors $c_{v,k}^{i,\gamma} = f^i_{v,\gamma} (v_k^{i,\gamma} ,v_k^{j,\gamma},\gamma_k^{i} ,\gamma_k^{j})$, $\forall j \in {\cal M}_i$.
The separation control signal $c_{d,k}^{i,\gamma}$ is applied to maintain a desired separation distance between the agents. This value relies on a function of the positions of each follower and its neighbors, such that
\begin{align}
  {c}_{d,k}^{i,\gamma}
  &
    = f^i_{d,\gamma} (\gamma_k^{i} ,\gamma_k^{j})
    = \frac{\sum_{j \in {\cal M}_i} {c}_{d,k}^{ij,\gamma}}{\abs{{\cal M}_i}},
    \label{sep}
\end{align}
where $\abs{{\cal M}_i}$ denotes the cardinality of ${\cal M}_i$ and ${c}_{d,k}^{ij,\gamma}$ represents the partial contribution of the separation control decision taken by agent~$i$ due to each agent~$j\in{\cal M}_i$.

The communication information between the agents are exchanged over a time-varying graph topology, where the position and velocity neighborhood measurements are available locally to each agent. Further, the measurements related to the leader are accessed by the flock members.
Due to the competing nature of the different objectives, due to their possibly conflicting nature, there is a compromise to be made so that the control action taken by each agent balances such conflicting goals. Hence, the control algorithm continuously updates the underlying strategies to improve the quality of attempted strategies.  

\section{Tracking Control Strategy}
\label{track}
The section introduces a model-free policy iteration process to compute the tracking control strategy ${c}_{t,k}^{i,\gamma}$ for each agent~$i$ to satisfy objective~\eqref{eq:objectives:track-leader}. This is accomplished in real-time and without knowing the dynamics of the leader or any of the followers.

\subsection{Optimization Framework}

The optimal control framework advises the tracking strategy while relying on tracking error measurements between the positions of the leader and agents. Each agent~$i$ uses an error vector where it stores the three recent tracking error measurements
$
\bm{Z}^{i,\gamma}_{k} = [{\gamma}^i_{k} - {\gamma}^\tau_{k},\ {\gamma}^i_{k-1} - {\gamma}^\tau_{k-1},\ {\gamma}^i_{k-2} - {\gamma}^\tau_{k-2} ]^T \in \mathbb{R}^3
$.
The number of error instances in that vector can be customized by the designer in order to balance between the required accuracy and complexity of the problem. Herein, a time window of three most recent error samples is found to be sufficient for the task in hand. The goal is to adapt the tracking strategies in order to annihilate the average tracking error $\epsilon_t$, which is defined as
\begin{equation*}
  \epsilon_t
  =
  \qty\Big( \sum_{i \in \{1,\ldots,M\} \setminus \{\tau\}} \norm{ \bm{g}_{k}^i - \bm{g}_{k}^{\tau} } )  \Big/  \qty\Big(M-1) - d_t
\end{equation*}
A performance index $\varsigma_0^{i,\gamma}=\sum^{\infty}_{k=0}W^{i,\gamma}_{k} ( \bm{Z}^{i,\gamma}_{k} ,c^{i,\gamma}_{t,k} )$ is considered to evaluate the quality of the tracking strategy at each time step for agent~$i$, where $W^{i,\gamma}_{k}$ is a convex quadratic utility function given by
\begin{equation}
  \label{Con}
  W^{i,\gamma}_{k} ( \bm{Z}^{i,\gamma}_{k} ,c^{i,\gamma}_{t,k} )
  = \dfrac{1}{2}
  \left[
    \bm{Z}^{i,\gamma T}_{k} \bm{J}^{i} \bm{Z}^{i,\gamma}_{k} +{K}^{i} ( c^{i,\gamma}_{t,k} )^{2}
  \right],
\end{equation}
where $\bm{0} < \bm{J}^i \in \mathbb{R}^{3 \times 3}$ and ${0} < K^i \in \mathbb{R}$ are weighting matrices for the tracking errors and control signal. The inequality symbols ``$>\bm{0}$'' and ``$\geq \bm{0}$'' refer to positive definite and positive semi-definite matrices, respectively.

The optimal control goal is to select the tracking strategy that minimizes the performance index $\varsigma^{i,\gamma}$ over the infinite horizon. First, a quadratic solving value function is assumed so that $ Q^{i,\gamma}( \bm{Z}^{i,\gamma}_{k} ,c^{i,\gamma}_{t,k}) \triangleq \varsigma_k^{i,\gamma}$. The structure of function $Q^{i,\gamma}$ is motivated by that of the utility function, such that
\begin{gather*}
  Q^{i,\gamma}( \bm{Z}^{i,\gamma}_{k} ,c^{i,\gamma}_{t,k} )
  =\frac{1}{2}
  \left[
    \begin{array} {cc}
      {\bm{Z}^{i,\gamma}_k}^T& {c^{i,\gamma}_{t,k}}
    \end{array}
  \right]
  \bm{G}^{i,\gamma}
  \left[
    \begin{array} {c}
      \bm{Z}^{i,\gamma}_k\\ c^{i,\gamma}_{t,k}
    \end{array}
  \right]
  \\
  \text{such that, }
  \bm{G}^{i,\gamma}
  \equiv
  \begin{bmatrix*}[l]
    \boldsymbol{G}^{i,\gamma}_{\boldsymbol{Z}^{i,\gamma}\boldsymbol{Z}^{i,\gamma}}
    & \boldsymbol{G}^{i,\gamma}_{\boldsymbol{Z}^{i,\gamma}{c_t^{i,\gamma}}}
    \\
    \boldsymbol{G}^{i,\gamma}_{{c_t^{i,\gamma}}\boldsymbol{Z}^{i,\gamma}}
    &
    \boldsymbol{G}^{i,\gamma}_{{c_t^{i,\gamma}}  {c_t^{i,\gamma}}}
  \end{bmatrix*}
  \in \mathbb{R}^{4\times4}
\end{gather*}
where $\bm{G}^{i,\gamma} > \bm{0}$, $\boldsymbol{G}^{i,\gamma}_{{c_t^{i,\gamma}}  {c_t^{i,\gamma}}} \in \mathbb{R}$, and $\boldsymbol{G}^{i,\gamma}_{{c_t^{i,\gamma}}\boldsymbol{Z}^{i,\gamma}} \in \mathbb{R}^{1 \times 3}$. 

This solving structure along with the infinite-horizon  performance index yield a temporal difference (Bellman) equation that is given by
\begin{equation}
  Q^{i,\gamma}( \bm{Z}^{i,\gamma}_{k} ,c^{i,\gamma}_{t,k})=W_k^{i,\gamma}( \bm{Z}^{i,\gamma}_{k} ,c^{i,\gamma}_{t,k} )+Q^{i,\gamma}( \bm{Z}^{i,\gamma}_{k+1} ,c^{i,\gamma}_{t,k+1} ).
  \label{Bell}
\end{equation}
Hence, Bellman's optimality condition is applied to find the optimal tracking strategy $c^{i,\gamma(*)}_{t,k} = \argmin_{c^{i,\gamma}_{t,k}}\left(Q^{i,\gamma}( \bm{Z}^{i,\gamma}_{k} ,c^{i,\gamma}_{t,k})\right)$. Thus,
\begin{equation}
  c^{i,\gamma(*)}_{t,k}
  =
  - \left({\boldsymbol{G}^{i,\gamma}_{{c_t^{i,\gamma}}  {c_t^{i,\gamma}}}}\right)^{-1} \boldsymbol{G}^{i,\gamma}_{{c_t^{i,\gamma}}\boldsymbol{Z}^{i,\gamma}} \boldsymbol{Z}^{i,\gamma}_{k}.
	\label{opt}
\end{equation}
Applying~\eqref{opt} into~\eqref{Bell} yields the Bellman optimality equation
\begin{multline}
  Q^{i,\gamma(*)}( \bm{Z}^{i,\gamma}_{k} ,c^{i,\gamma(*)}_{t,k})
  =
  W_k^{i,\gamma}( \bm{Z}^{i,\gamma}_{k} ,c^{i,\gamma(*)}_{t,k}) \\
  + Q^{i,\gamma(*)}( \bm{Z}^{i,\gamma}_{k+1} ,c^{i,\gamma(*)}_{t,k+1} ).
  \label{Bello}
\end{multline}
Solving~\eqref{opt} and~\eqref{Bello} simultaneously for each agent would solve the underlying guidance or optimal trajectory tracking problem. Therefore, approximate or regression methods are needed to implement the solutions of~\eqref{opt} and~\eqref{Bello} in real-time.

A RL technique based on Policy Iteration (PI) is adopted to provide an online solution for this problem.
This is done recursively by solving the following temporal difference (Bellman) form:
\begin{multline*}
  Q^{i,\gamma(r)}( \bm{Z}^{i,\gamma}_{k} ,c^{i,\gamma}_{t,k}) - Q^{i,\gamma(r)}( \bm{Z}^{i,\gamma}_{k+1} ,c^{i,\gamma}_{t,k+1} )
  = \\  W_k^{i,\gamma(r)}( \bm{Z}^{i,\gamma}_{k} ,c^{i,\gamma}_{t,k}).
\end{multline*}
Then, the policy is updated using
\begin{equation}
  c^{i,\gamma(r+1)}_{t,k}
  = -\left( \left({\boldsymbol{G}^{i,\gamma}_{{c_t^{i,\gamma}}  {c_t^{i,\gamma}}}}\right)^{-1} \boldsymbol{G}^{i,\gamma}_{{c_t^{i,\gamma}}\boldsymbol{Z}^{i,\gamma}}\right)^{(r)} \boldsymbol{Z}^{i,\gamma}_{k}.
  \label{RLS_Pol}
\end{equation}
This solution algorithmic form is implemented using a recursive least squares regression approach, which is explained below. 

\subsection{Recursive Least Squares}
The PI solution is implemented in two steps. First, a given policy is evaluated. Second, the tracking strategy-to-follow is improved. Therefore, the RLS approach solves~\eqref{Bell} for the optimal value $\bm{G}^{i,\gamma}$ or strategy using value function approximation since it is not possible to solve Bellman optimality equation analytically.
The approximated value function $\tilde Q^{i,\gamma}$ is represented by
\begin{gather*}
  \tilde Q^{i,\gamma}( \bm{Z}^{i,\gamma}_{k} ,\tilde c^{i,\gamma}_{t,k} )
  =
  \frac{1}{2}
  \left[
    \begin{array} {cc}
      \bm{Z}^{{i,\gamma}^T}_k  & \tilde c^{i,\gamma}_{t,k}
    \end{array}
  \right]
  \bm{\Psi}^{i,\gamma}
  \left[
    \begin{array} {c}
      \bm{Z}^{i,\gamma}_k\\ \tilde c^{i,\gamma}_{t,k}
    \end{array}
  \right]
  \\
  \text{such that, }
  \bm{\Psi}^{i,\gamma}
  \equiv
  \begin{bmatrix*}[l]
    \boldsymbol{\Psi}^{i,\gamma}_{\boldsymbol{Z}^{i,\gamma}\boldsymbol{Z}^{i,\gamma}}
    & \boldsymbol{\Psi}^{i,\gamma}_{\boldsymbol{Z}^{i,\gamma}{\tilde c_t^{i,\gamma}}}
    \\
    \boldsymbol{\Psi}^{i,\gamma}_{{\tilde c_t^{i,\gamma}}\boldsymbol{Z}^{i,\gamma}}
    &
    \boldsymbol{\Psi}^{i,\gamma}_{{\tilde c_t^{i,\gamma}}  {\tilde c_t^{i,\gamma}}}
  \end{bmatrix*}
  \in \mathbb{R}^{4\times4},
\end{gather*}
where $\bm{\Psi}^{i,\gamma} > 0$, $\boldsymbol{\Psi}^{i,\gamma}_{{\tilde c_t^{i,\gamma}}  {\tilde c_t^{i,\gamma}}} \in \mathbb{R}$, and $\boldsymbol{\Psi}^{i,\gamma}_{{\tilde c_t^{i,\gamma}}\boldsymbol{Z}^{i,\gamma}} \in \mathbb{R}^{1 \times 3}$.

Let
$\bm{\varphi}^{i,\gamma}_{k} = \bm{\bar x}^{i,\gamma}_{k} - \bm{\bar x}^{i,\gamma}_{k+1}$, where
$\bm{\bar x}^{i,\gamma} = [ 0.5z_1^2$  $z_1z_2 \ z_1z_3 \ z_1z_4 \ 0.5z_2^2 \ z_2z_3 \ z_2z_4 \ 0.5z_3^2 \ z_3z_4$ $0.5z_4^2 ]$
and $z_1$ to $z_4$ correspond to the elements of vector $\vect{ \bm{Z}^{{i,\gamma}^T}  , \tilde c^{i,\gamma}_{t} }$.
Also, let $\bm{\vartheta}^{i,\gamma}\  \in\mathbb{R}^{10\times 1}$ be the entries in the symmetric solution matrix $\bm{\Psi}^{i,\gamma}$ associated with the vector $\bm{\bar x}^{i,\gamma}$. Hence, it is required to solve the following equation
\begin{equation}
\bm{\varphi}^{i,\gamma}_{k}\bm{\vartheta}^{i,\gamma}=W_k^{i,\gamma}.
\end{equation}
This is done using RLS technique for each agent~$i$.
Then, the approximated optimal strategy-to-follow is calculated by~\eqref{RLS_Pol} after reconstructing $\bm{\Psi}^{i,\gamma}$ back from $\bm{\vartheta}^{i,\gamma}$, such that the improved control strategy follows
$
\tilde c^{i,\gamma}_{t,k}=- \left({\boldsymbol{\Psi}^{i,\gamma}_{{\tilde c_t^{i,\gamma}}  {\tilde c_t^{i,\gamma}}}}\right)^{-1} \boldsymbol{\Psi}^{i,\gamma}_{{\tilde c_t^{i,\gamma}}\boldsymbol{Z}^{i,\gamma}} \boldsymbol{Z}^{i,\gamma}_{k}
$.
The RLS approach solves for the unknown weights $\bm{\vartheta}^{i,\gamma}_{k}  \in\mathbb{R}^{10\times 1}$  in real-time as follows~\cite{aastrom2013adaptive}, without requiring any prior knowledge about the agents' dynamics
\begin{align*}
  {\bm{\vartheta}}^{i,\gamma}_k & = {\bm{\vartheta}}^{i,\gamma}_{k-1}+ \bm L^{i,\gamma}_k \left(W_k^{i,\gamma}-\bm{\varphi}_k^{{i,\gamma}} {\bm{\vartheta}}^{i,\gamma}_{k-1} \right) 
  \\
  \bm{L}_k^{i,\gamma} & = \bm{P}_{k-1}^{i,\gamma} \bm{\varphi}_k^{{i,\gamma}^T} \left(1 + \bm{\varphi}^{{i,\gamma}}_k \bm{P}^{i,\gamma}_{k-1} \bm{\varphi}^{{i,\gamma}^T}_k \right)^{-1} 
  \\
  \bm{P}^{i,\gamma}_k & = \left(\bm{I} - \bm{L}^{{i,\gamma}}_k\bm{\varphi}^{i,\gamma}_k \right) \bm{P}^{i,\gamma}_{k-1} 
\end{align*}
where $\bm{L}^{{i,\gamma}} \in \mathbb{R}^{10 \times 1}$ is a gain adaptation vector, $\bm{P}^{{i,\gamma}} \in \mathbb{R}^{10 \times 10}$ is a covariance matrix, and $\bm{I}$ is the identity matrix.
A PI solution using the RLS method is shown in \Cref{alg:alg1}. It is executed simultaneously by each agent~$i$. 

\begin{algorithm}[htb]
  \setstretch{1} 
  \caption{RLS-Based PI Tracking Mechanism}\label{alg:alg1}
  \begin{algorithmic}[1] 
    \Require
    \Statex Weighting matrices $\boldsymbol{J}^{i}$ and ${K}^{i}$
    \Statex Number of search iterations ${\cal T}_n$
    \Statex Initial tracking error vector $\bm{Z}_0^{i,\zeta}$
    \Statex Initial weights $\bm{\Psi}^{i,\gamma(0)}$ and the corresponding ${\bm{\vartheta}}^{i,\gamma}_0$
    \Statex Initial $\bm{P}^{i,\gamma}_0$ and $\bm{L}^{i,\gamma}_0$
    \Statex Convergence error $\xi$ and size of a moving window $\cal W$
    \Ensure
    \Statex Tuned weights $\bm{\Psi}^{i,\gamma(*)}$
    \Statex
    \State $k \gets 0$ 
    \State RLS\_weights\_converged $\gets$ false
    \While {$k <{\cal T}_n$ \AlgAnd RSL\_weights\_converged $=$ false}
    \State Compute control signal $\tilde c^{i,\gamma}_{t,k}$ and apply it to agent~$i$
    \State Find $W_k^{i,\gamma}( \bm{Z}^{i,\gamma}_{k} ,\tilde c^{i,\gamma}_{t,k})$ and $\bm{\bar x}^{i,\gamma}_{k}$
    \State Calculate $\bm{Z}^{i,\zeta}_{(k+1)} ,\tilde c^{i,\gamma}_{t,k+1}$
    \State Find $\bm{\bar x}^{i,\gamma}_{k+1}$ and calculate $ \bm{\varphi}^{i,\gamma}_{k}$
    \State $k \gets k+1$ 
    \State Calculate $\bm{P}^{i,\gamma}_k$ and $\bm{L}^{i,\gamma}_k$
    \State Find ${\bm{\vartheta}}^{i,\gamma}_k$, then update $\bm{\Psi}^{i,\gamma}_k$
    \If{$k > {\cal W}$ \AlgAnd $\norm{\bm{\vartheta}^{i,\zeta \, (k+1-l)}-\bm{\vartheta}^{i,\zeta(k-l)}} \le \xi,$ $\forall l \in \{0,1,\ldots,{\cal W}\}$,}
    \State $\bm{\Psi}^{i,\gamma(*)} \gets \bm{\Psi}^{i,\gamma(k+1)}$ 
    \State RSL\_weights\_converged $\gets$ true
    \EndIf
    \EndWhile
    \State \Return $\bm{\Psi}^{i,\gamma(*)}$
  \end{algorithmic}
\end{algorithm}

\section{Consensus Control Strategy}
\label{cons}
The second objective is to achieve cohesiveness among the moving agents through a consensus protocol, as expressed in~\eqref{eq:objectives:velocity-consensus}. This is done using the means of a time-varying communication graph topology  $\mathcal{G} = \{ \mathcal{M} , \mathcal{L} \}$ , where $\mathcal{M} = \{\sigma_i\}_{i=1,\ldots,\abs{\mathcal{M}}}$ is the set of graph nodes of cardinality $\abs{\mathcal{M}}$, and $\mathcal{L} = \{ (\sigma_i,\sigma_j) \in \mathcal{M}^2 \}$  is the set of undirected edges~\cite{casteigts2010}. The connection strength of each edge $(\sigma_i,\sigma_j) \in \mathcal{L}$ in the undirected graph is denoted as $s_{ij}=s_{ji}$, for $j\neq i$, where $s_{ii}=0$. The consensus control strategy of each agent~$i$ is calculated using
\begin{equation}
  {c}^{i,\gamma}_{v,k} = - \sum _{j \in \mathcal{M}_{i}} s_{ij}(v^{i,\gamma}_{k} - v^{j,\gamma}_{k}).
  \label{Consensus}
\end{equation}
The graph connectivity weights are decided using a scalar pump function ${\delta _a}(\gamma ^{ij})$, such that	
\begin{equation*}
  {\delta _a}(\gamma ^{ij})= 
  \begin{cases}
    1 &, \gamma ^{ij} \in [0,a) \\
    \frac{1}{2}\left[1+\cos(\pi \frac{\gamma ^{ij}-a}{r-a})\right] 
    &, \gamma ^{ij} \in [a,r)   \\
    0 &, \text{otherwise}    
  \end{cases}
\end{equation*}
where $\gamma ^{ij} = \gamma ^{i} -\gamma ^{j}$ is the relative distance between agents $i$ and $j$. Note that the output of the scalar pump function also takes into account a connectivity communication range $r$ between the agents, which somewhat contributes to the separation control objective~\cite{Olfati2006}.
The edge or connectivity weights are calculated as
\begin{equation*}
  s_{ij}(\gamma^{ij}) = {\delta _a}\left(\norm{\gamma^{ij}}_\alpha/\norm{r}_\alpha \right) \in [0,1), \ j \neq i.
\end{equation*} 
The $\alpha$-norm is defined by
\begin{equation*}
  \norm{\gamma^{ij}}_\alpha = \frac{1}{\mu} \left[\sqrt{1+\mu\norm{\gamma^{ij}}^2}-1 \right],
\end{equation*}
where $\mu$ is a positive real scalar.
This control strategy is adaptive to the agents formation, where the connectivity between the agents can vary in time according to the agents proximity to one other.

\section{Separation Control Strategy}
\label{separation}

The separation control strategy~\eqref{eq:objectives:separation} prevents agents from colliding by enforcing repulsive-attraction forces in order to control the localization of agents with respect to each other. A Fuzzy RL adaptation mechanism that uses a zero-order Tagaki-Sugeno (TS) fuzzy logic inference system which is implemented using an online value iteration is considered herein, as detailed in~\cite{ICRA21}. The RL approach adapts the consequences of the fuzzy rules in real-time. This works according to a temporal difference reward that penalizes the agents form getting closer to each other and vice versa. 
The approach aggregates the separation policies or decisions made for each agent~$i$ in reaction to the other agents in its neighborhood ${\cal M}_i$, which results in an overall signal  ${c}_{d,k}^{i,\gamma}$ for each agent~$i$ in the $\gamma$-direction (recall that $\gamma \in \{x,y\}$), such that
\begin{align}
  {c}_{d,k}^{i,\gamma}
  &
    = \frac{\displaystyle \sum_{j \in {\cal M}_i} \displaystyle \sum_{f=1}^{\cal F} 
    \Theta_k^{ij,\gamma (f)}  \eta^{ij,\gamma(f)}}{\abs{{\cal M}_i}},
    \label{sepo}
\end{align}
where $\cal F$ is the total number of fuzzy rules. For each rule $f\in\{1,\ldots,{\cal F}\}$, $\Theta_k^{ij,\gamma (f)}$ is the firing strength and $\eta^{ij,\gamma(f)}$ is the consequence of that rule~$f$. The latter is tuned online using the RL process presented in~\cite{ICRA21}.

\section{Results}
\label{Sim}

\begin{figure*}[h!]
  \centering
  \subcaptionbox{%
    \label{fig:video-scenes-1}}
  {%
    \includegraphics[width=0.33\textwidth]{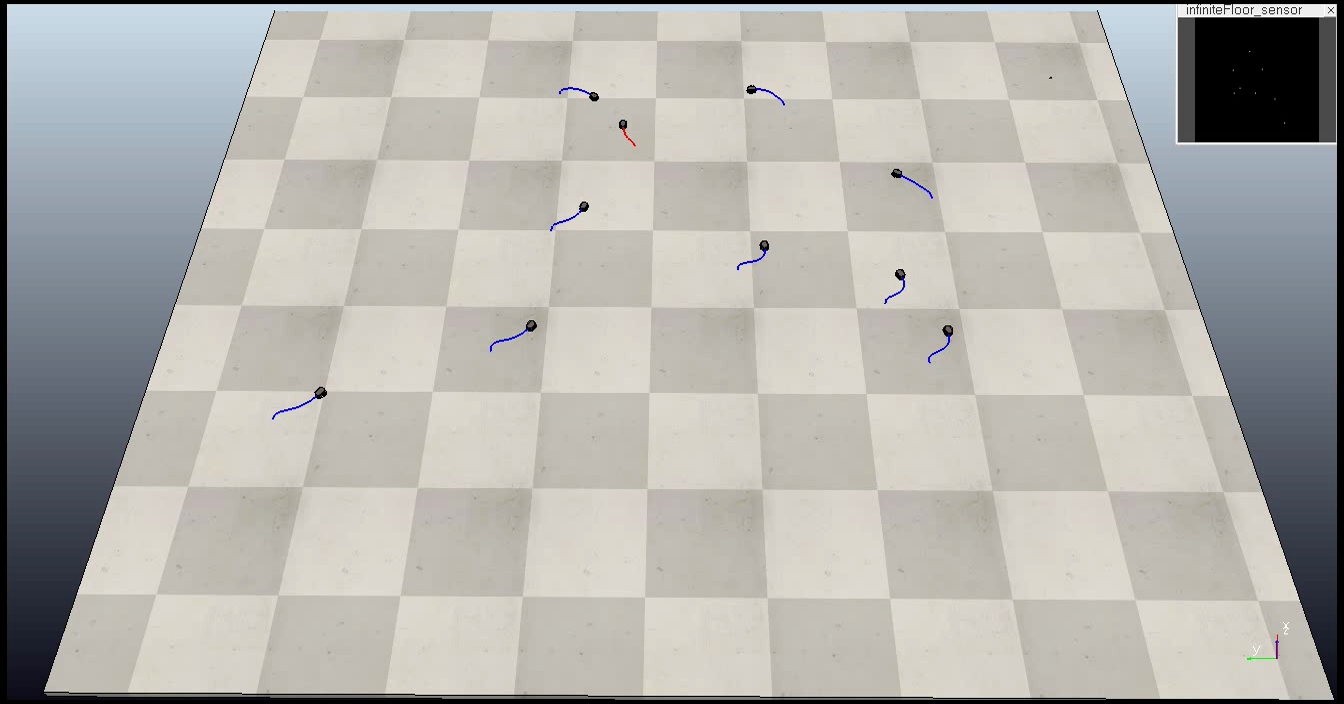}%
  }%
  \hfill
  \subcaptionbox{%
    \label{fig:video-scenes-2}}
  {%
    \includegraphics[width=0.33\textwidth]{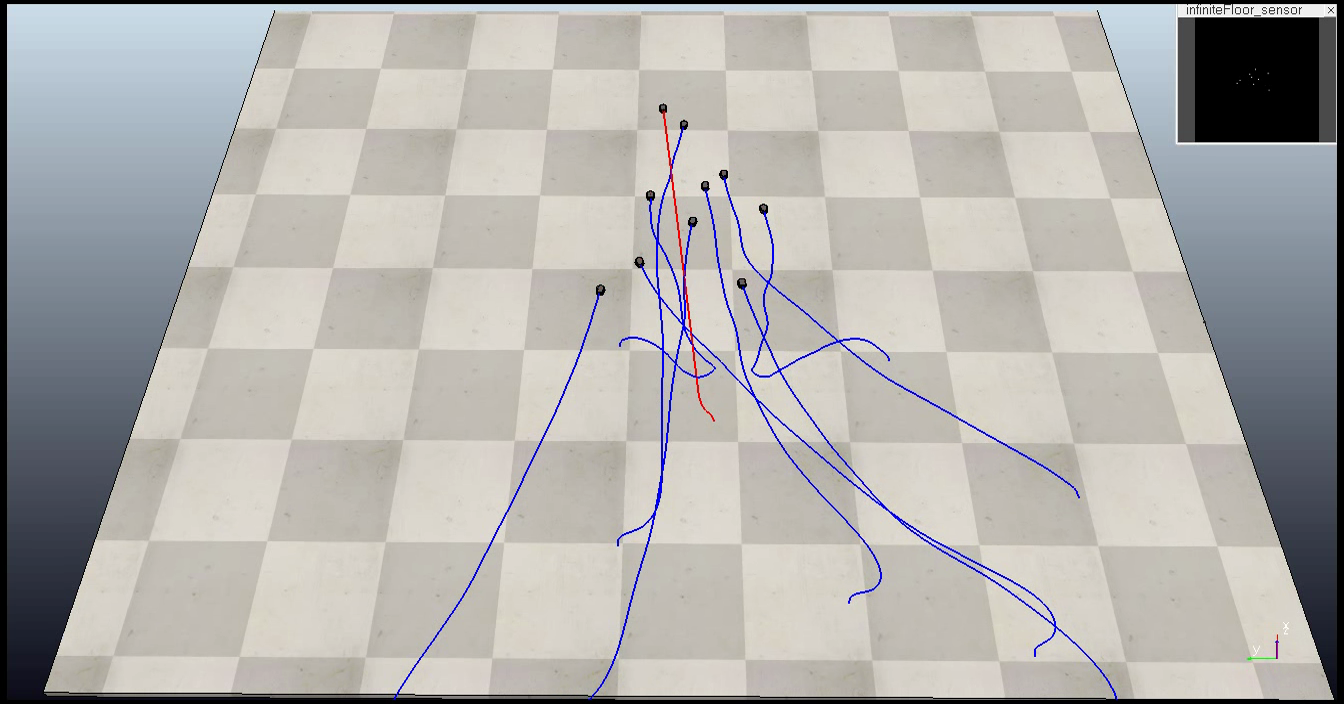}%
  }%
  \hfill
  \subcaptionbox{%
    \label{fig:video-scenes-3}}
  {%
    \includegraphics[width=0.33\textwidth]{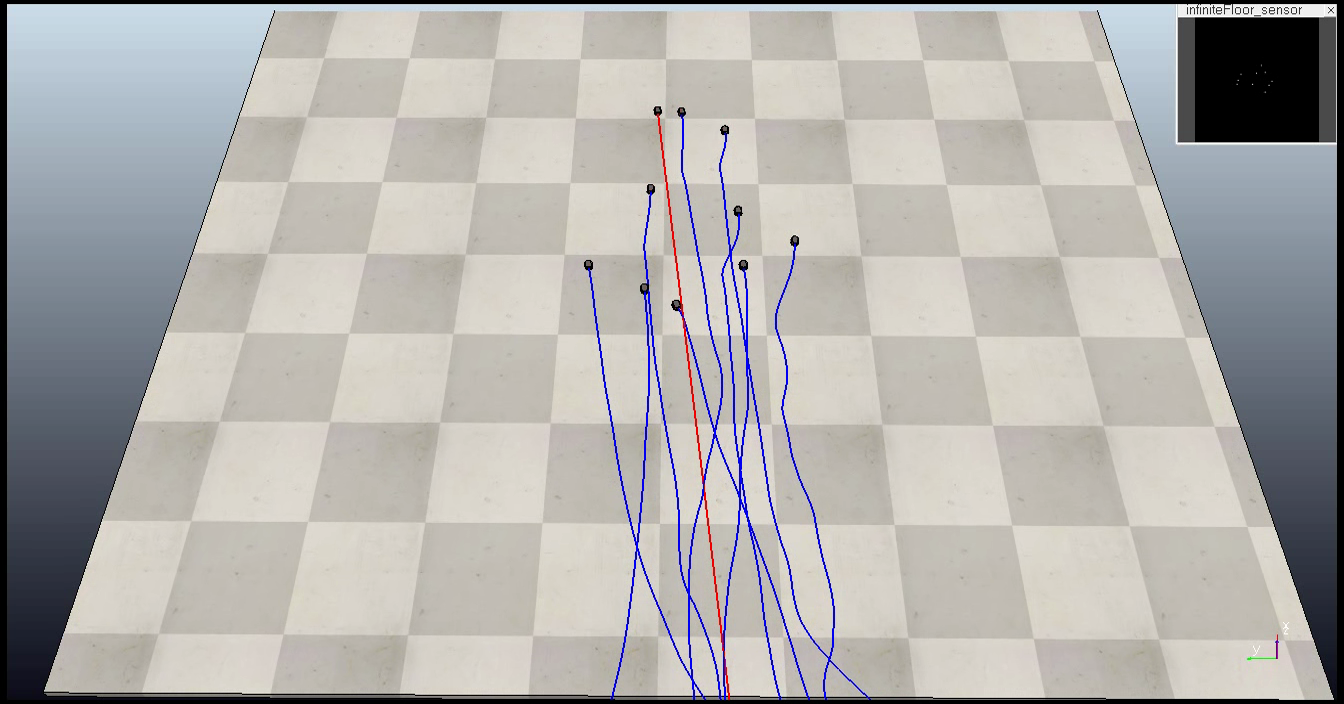}%
  }%
  \\
  %
  %
  \subcaptionbox{%
    \label{fig:video-scenes-4}}
  {%
    \includegraphics[width=0.33\textwidth]{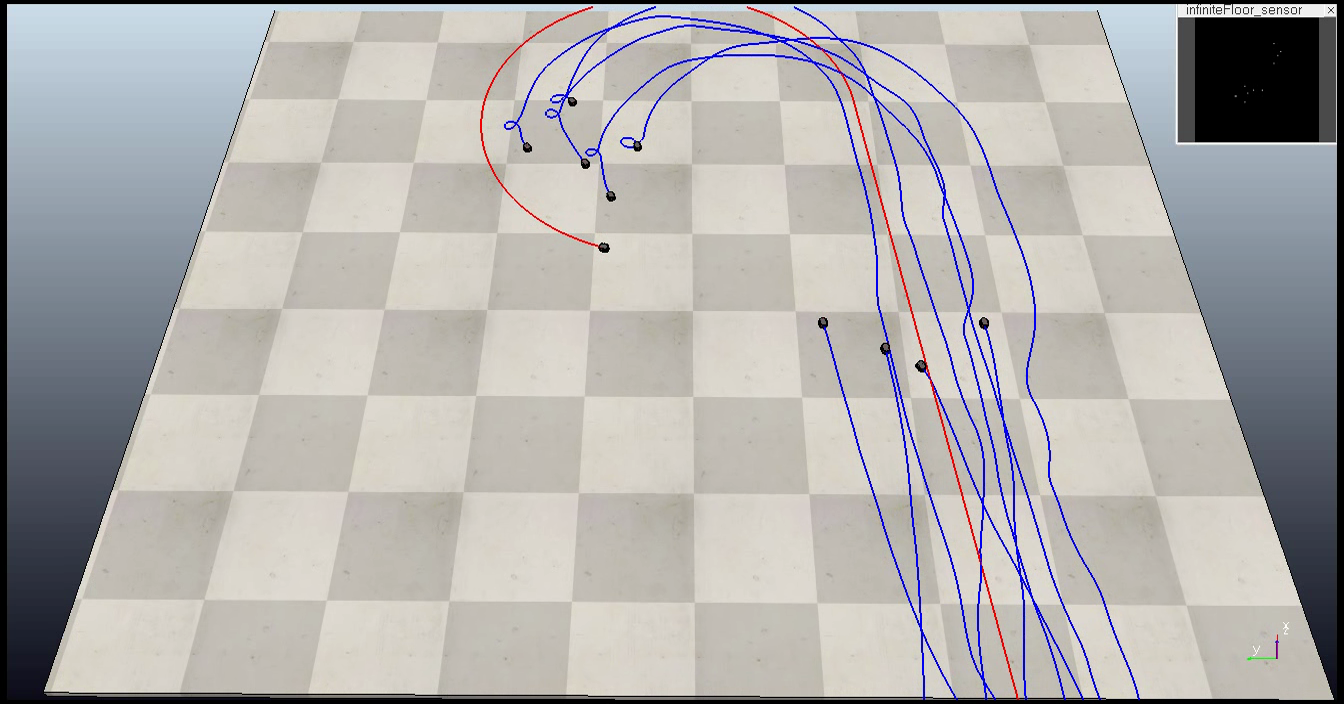}%
  }%
  \hfill
  \subcaptionbox{%
    \label{fig:video-scenes-5}}
  {%
    \includegraphics[width=0.33\textwidth]{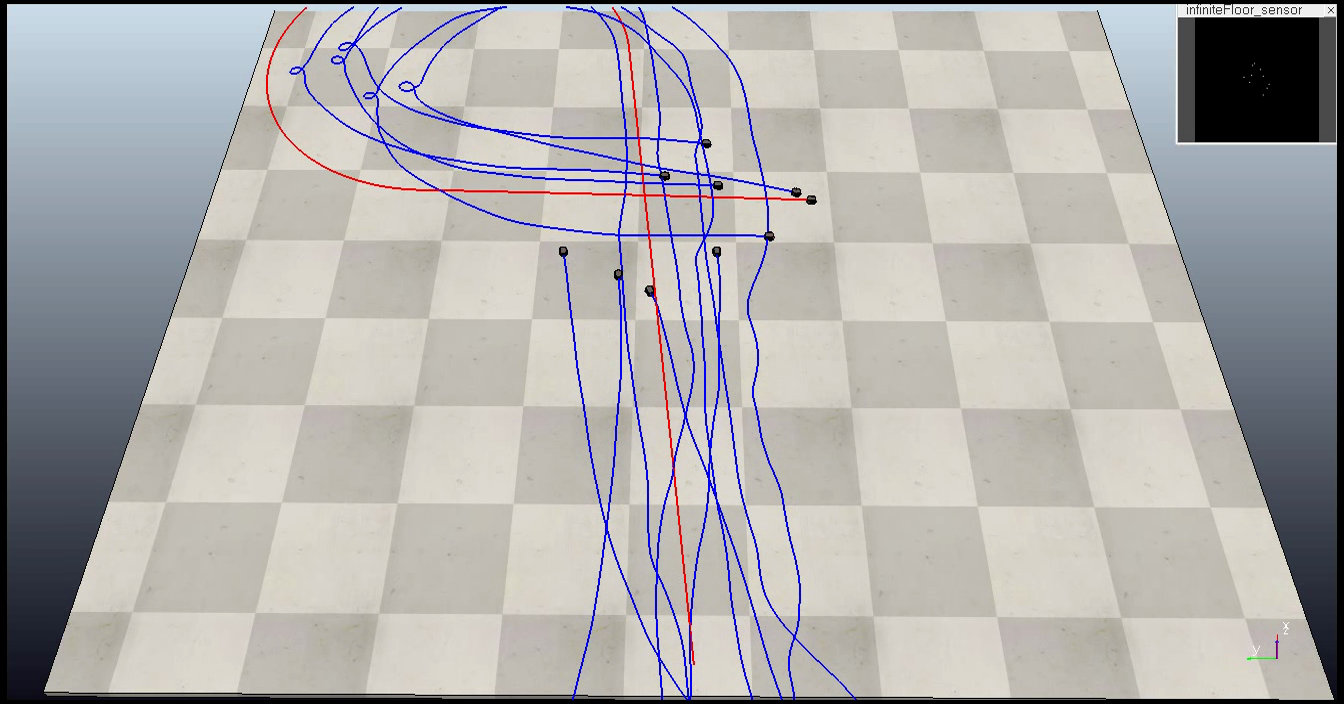}%
  }%
  \hfill
  \subcaptionbox{%
    \label{fig:video-scenes-6}}
  {%
    \includegraphics[width=0.33\textwidth]{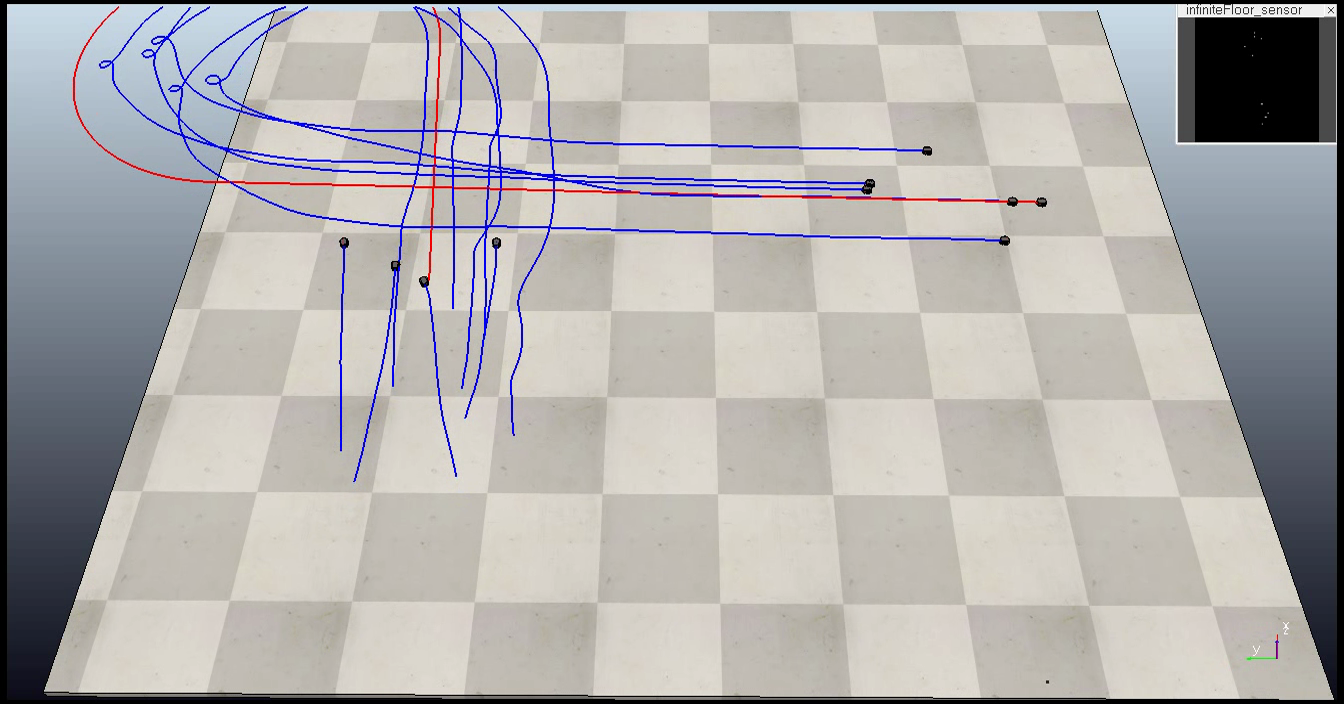}%
  }%
  \caption{Flock trajectory\label{fig:video-scenes}} 
\end{figure*}

A system of 10 Pioneer-3DX{\texttrademark} mobile robots is simulated to validate the proposed policy iteration approach in \Cref{alg:alg1}, where one robot plays the role of a leader while the rest of the robots act as followers. The simulation is realized in CoppeliaSim{\texttrademark}\footnote{\url{https://coppeliarobotics.com}~[accessed: July 10, 2021]}, a realistic robot simulation software that adopts some of the state-of-the-art physics engines to simulate physical phenomena, such as gravity, friction, etc. The robots maximum linear velocity and acceleration are set to $\SI{1.2}{\m/\s}$ and $\SI{2}{\m/\s^2}$, respectively, while the angular velocity and acceleration are capped at $\pm\SI{150}{\degree/\s}$ and $\pm\SI{150}{\degree/\s^2}$, respectively. Initially, the robots are scattered randomly in the environment, as seen in \Cref{fig:video-scenes-1}. The trajectory of the flock is shown in \Cref{fig:video-scenes}, where the leader's trajectory is marked in red. By default, the desired separation distance between agents and the average tracking proximity are taken as $s=\SI{2}{\m}$ and $d_t=s$. During the \SI{65}{\s} simulation, the leader assumes different types of trajectories as shown in \Cref{tab:leader-velocities}.
\begin{table}
  \centering
  \caption{Leader velocity}
  \label{tab:leader-velocities}
  \sisetup{
      round-mode        = places,  
      round-precision   = 1, 
    }
  \begin{tabular}{c|S|S}
    \toprule
    {Time period}   &  {Linear velocity}  &  {Angular velocity}\\
    {[\si{\s}]}     &  {[\si{\m/\s}]}     &  {[\si{\degree/\s}]}\\
    \midrule
    $[0,20]$   & 0.9 & 0\\
    $[20,25]$  & 0.9 & 0\\
    $[25,45]$  & 1.2 & 6.6\\
    $[45,65]$  & 1.2 & 0\\
    \bottomrule
  \end{tabular}
\end{table}
To further challenge the controller, 4 followers are decommisioned at the \SI{20}{s} milestone, as can be seen in \Cref{fig:video-scenes-4,fig:video-scenes-5,fig:video-scenes-6}, and the desired separation distance is increased to $s=\SI{2.5}{\m}$ at time \SI{45}{\s}. In other words, there are four stages in the simulation, distributed as $[ 0 \ldots \SI{20}{\s} \ldots \SI{25}{\s} \ldots \SI{45}{\s} \ldots \SI{65}{\s} ]$.
The weighting matrices of the utility function are set to $\bm{J}^{i}=0.0001\times \bm{I}_{3\times3}$ and ${K}^{i}=0.01$, $\forall i$. The RLS simulation parameters are taken as ${\cal T}_n=400$ and $\bm{P}^{i,\gamma}_0=100\times \bm{I}_{10\time 10}$, $\forall i$. The parameters of the position dependent adjacency function are fixed to $a=1$, $r=\SI{3.5}{\m}$, and $\mu = 0.5$.

\begin{figure*}[h!]
  \centering
  \subcaptionbox{Average tracking error~$\epsilon_t$%
    \label{fig:PI-vs-VI-average-tracking-error}}
  {%
    \includegraphics[width=0.33\textwidth]{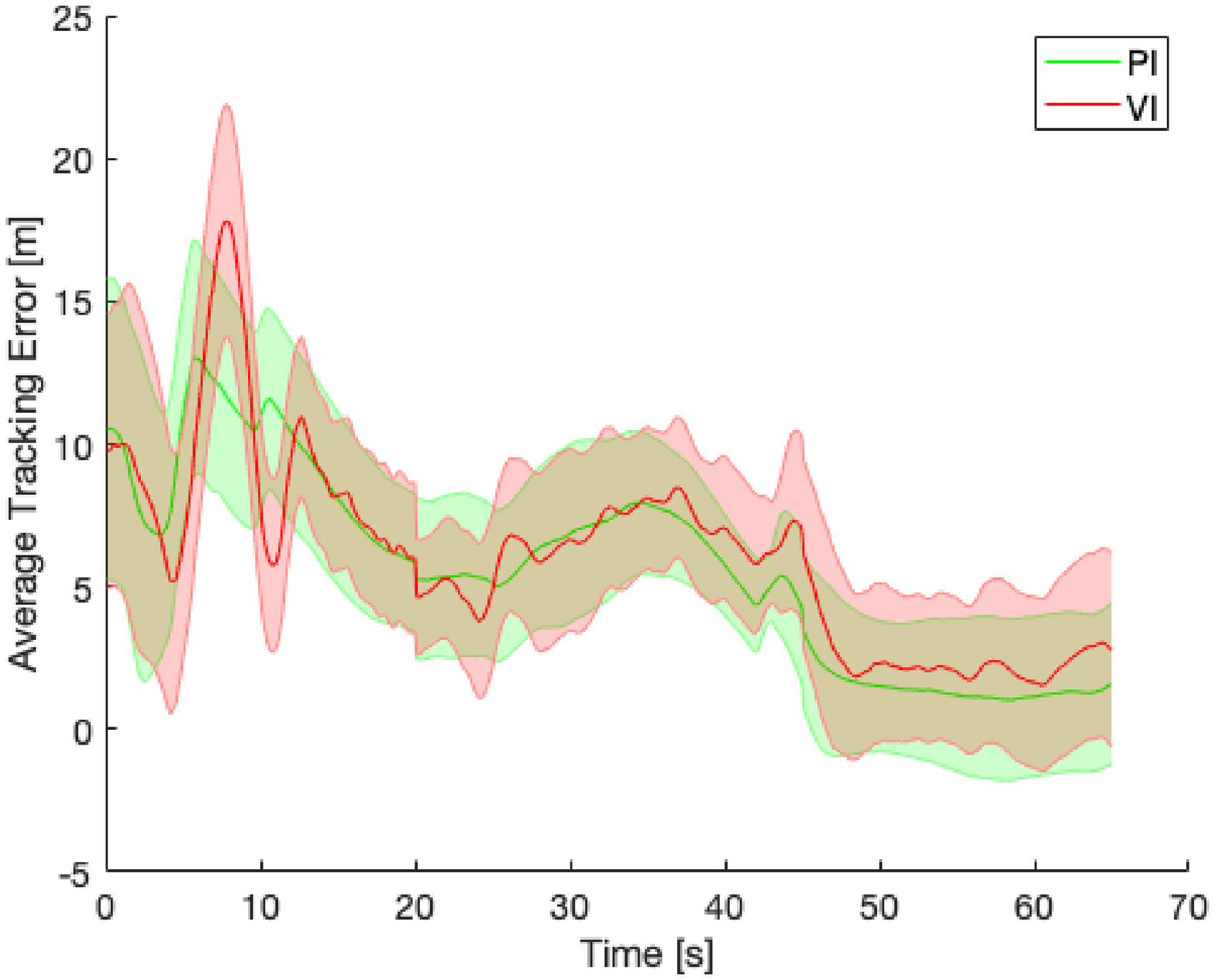}%
  }%
  \hfill
  \subcaptionbox{Average separation error%
    \label{fig:PI-vs-VI-average-separation-error}}
  {%
    \includegraphics[width=0.33\textwidth]{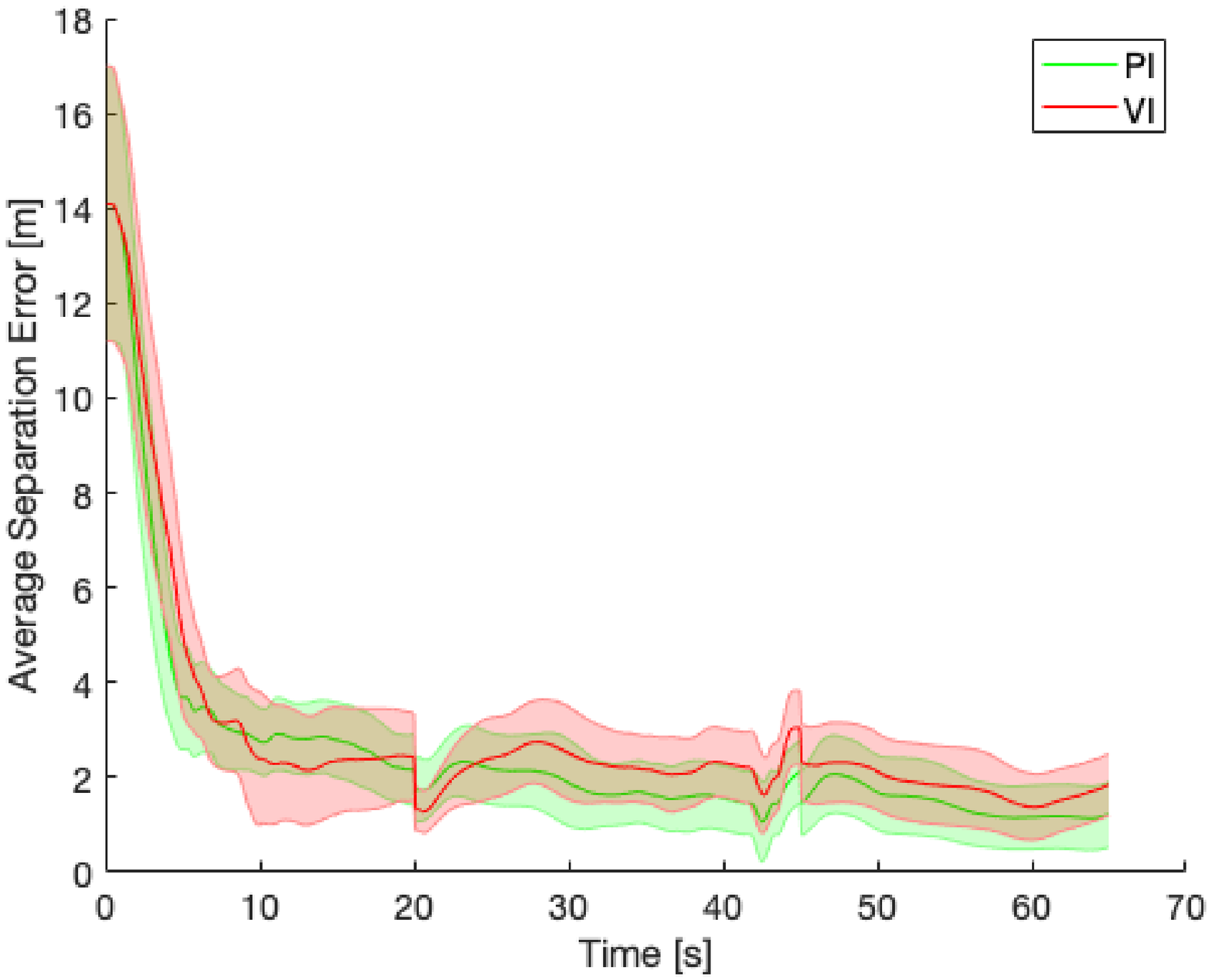}%
  }%
  \hfill
  \subcaptionbox{Average follower velocity%
    \label{fig:PI-vs-VI-average-follower-velocity}}
  {%
    \includegraphics[width=0.33\textwidth]{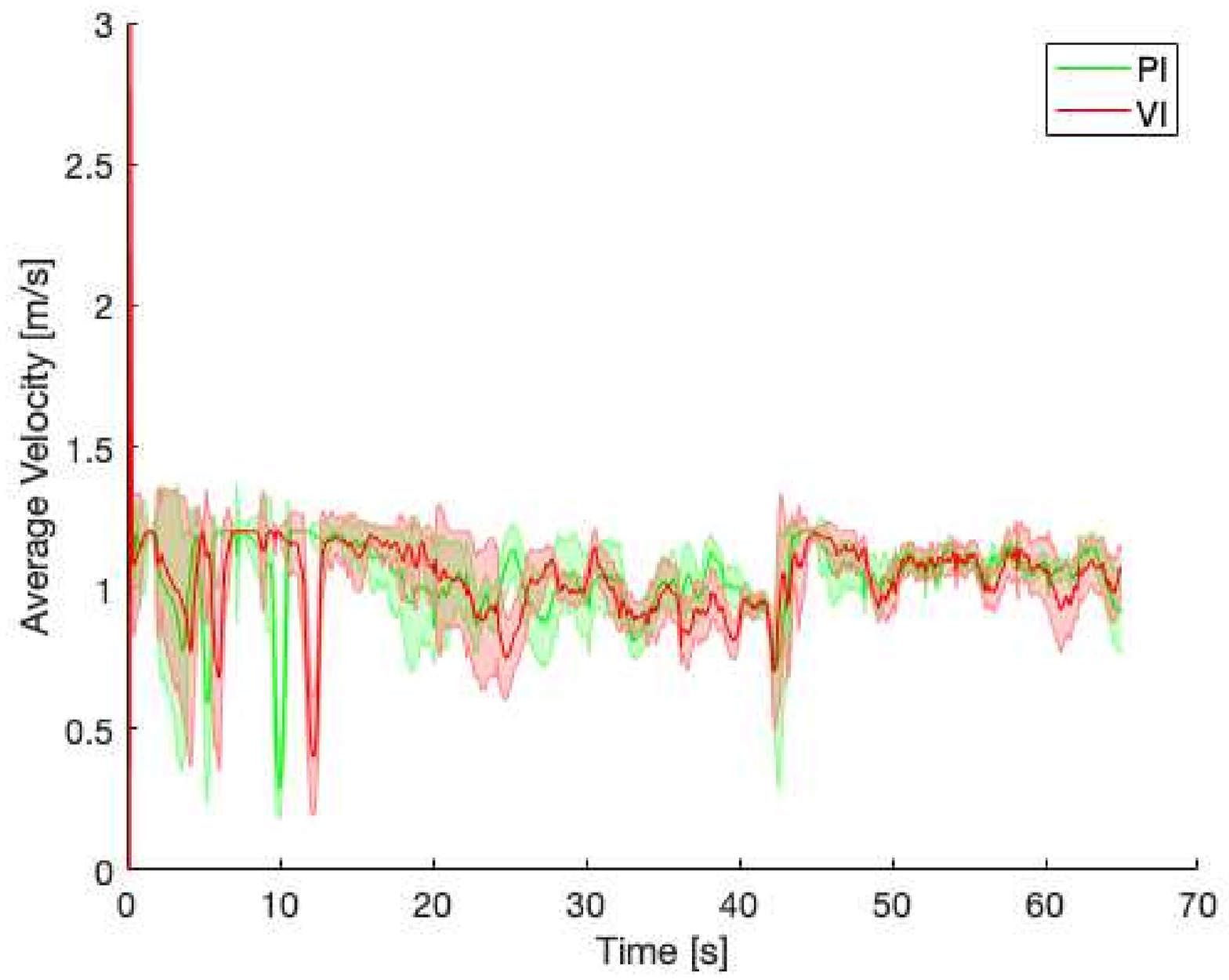}%
  }%
  \caption{Comparison with the Value Iteration approach of~\cite{ICRA21}\label{fig:PI-vs-VI}} 
\end{figure*}

For a more quantitative assessment of the proposed control algorithm, its performance is contrasted against the Fuzzy-RL Value Iteration approach introduced in~\cite{ICRA21}. The results are shown in \Cref{fig:PI-vs-VI}, where shaded areas indicate the standard deviation. It is clear from \Cref{fig:PI-vs-VI-average-follower-velocity} that the followers' average velocity stabilizes in each phase of the simulation, and that in this aspect, the performance of the PI approach is relatively close to that of the VI. As for the average tracking error, it is noticed that it is generally smaller with the PI techniques, as revealed in \Cref{fig:PI-vs-VI-average-tracking-error}. One can also observe how it is faster to converge towards the end of the simulation than with the VI method. The same remark is applicable to the average separation error between the followers and their neighbors, with the addition that the PI algorithm demonstrates a higher robustness here represented by the smaller standard deviation, as shown in \Cref{fig:PI-vs-VI-average-separation-error}.

\begin{figure*}[h!]
  \centering
  \subcaptionbox{PI ($x$)%
    \label{fig:tracking-critic-weights-PI-x}}
  {%
    \includegraphics[width=0.49\columnwidth]{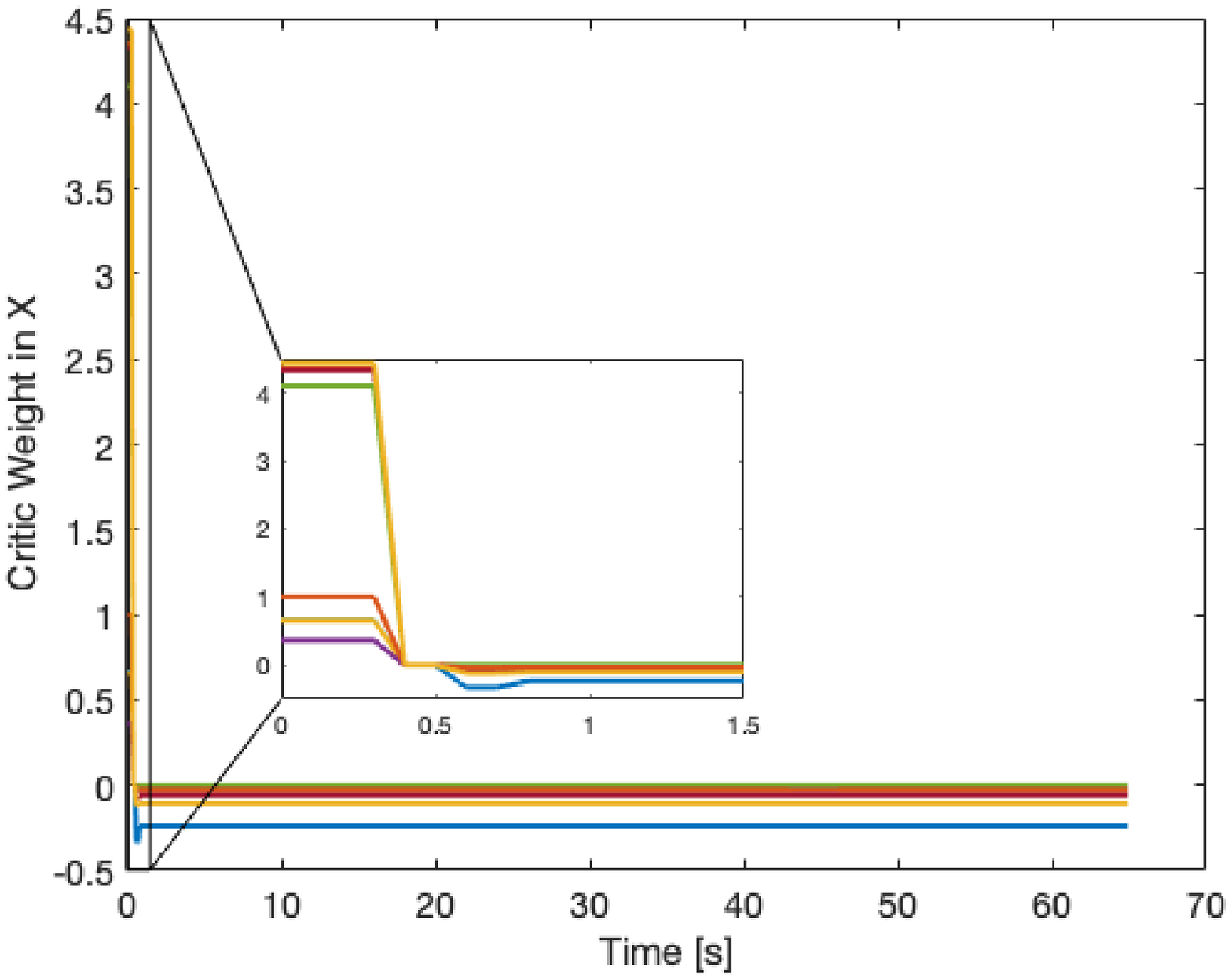}%
  }%
  \hfill
  \subcaptionbox{PI ($y$)%
    \label{fig:tracking-critic-weights-PI-y}}
  {%
    \includegraphics[width=0.49\columnwidth]{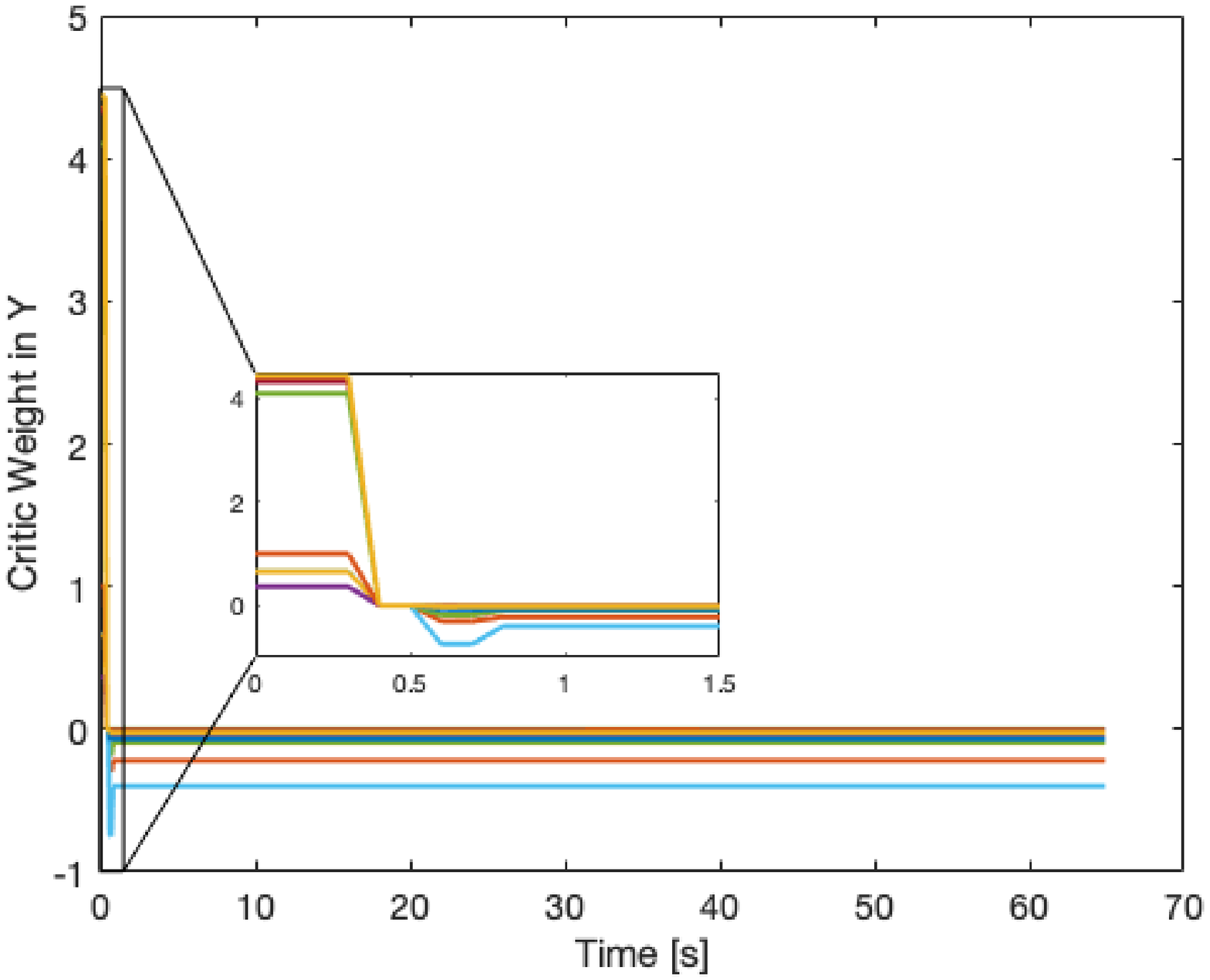}%
  }%
  \hfill
  \subcaptionbox{VI ($x$)%
    \label{fig:tracking-critic-weights-VI-x}}
  {%
    \includegraphics[width=0.49\columnwidth]{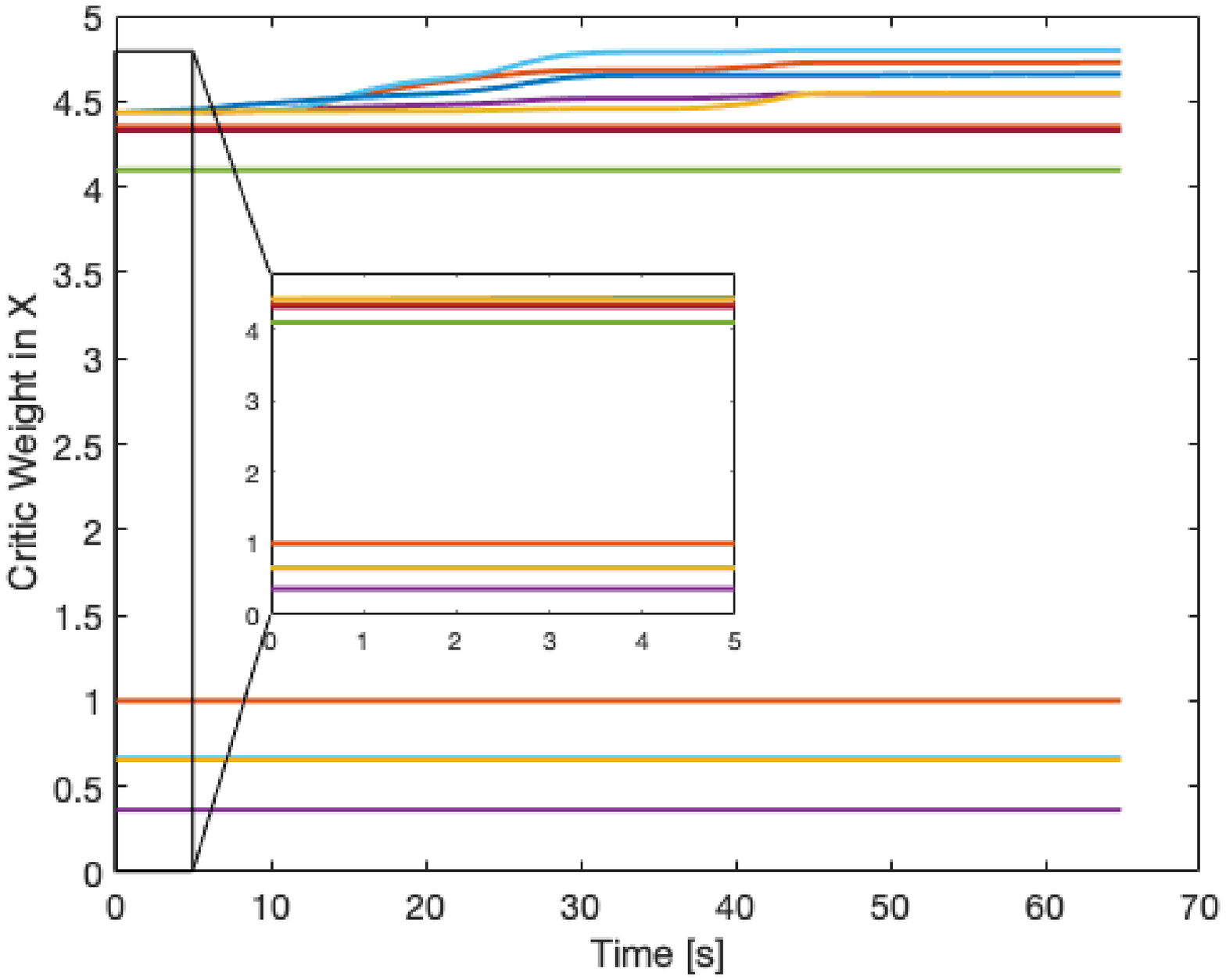}%
  }%
  \hfill
  \subcaptionbox{VI ($y$)%
    \label{fig:tracking-critic-weights-VI-y}}
  {%
    \includegraphics[width=0.49\columnwidth]{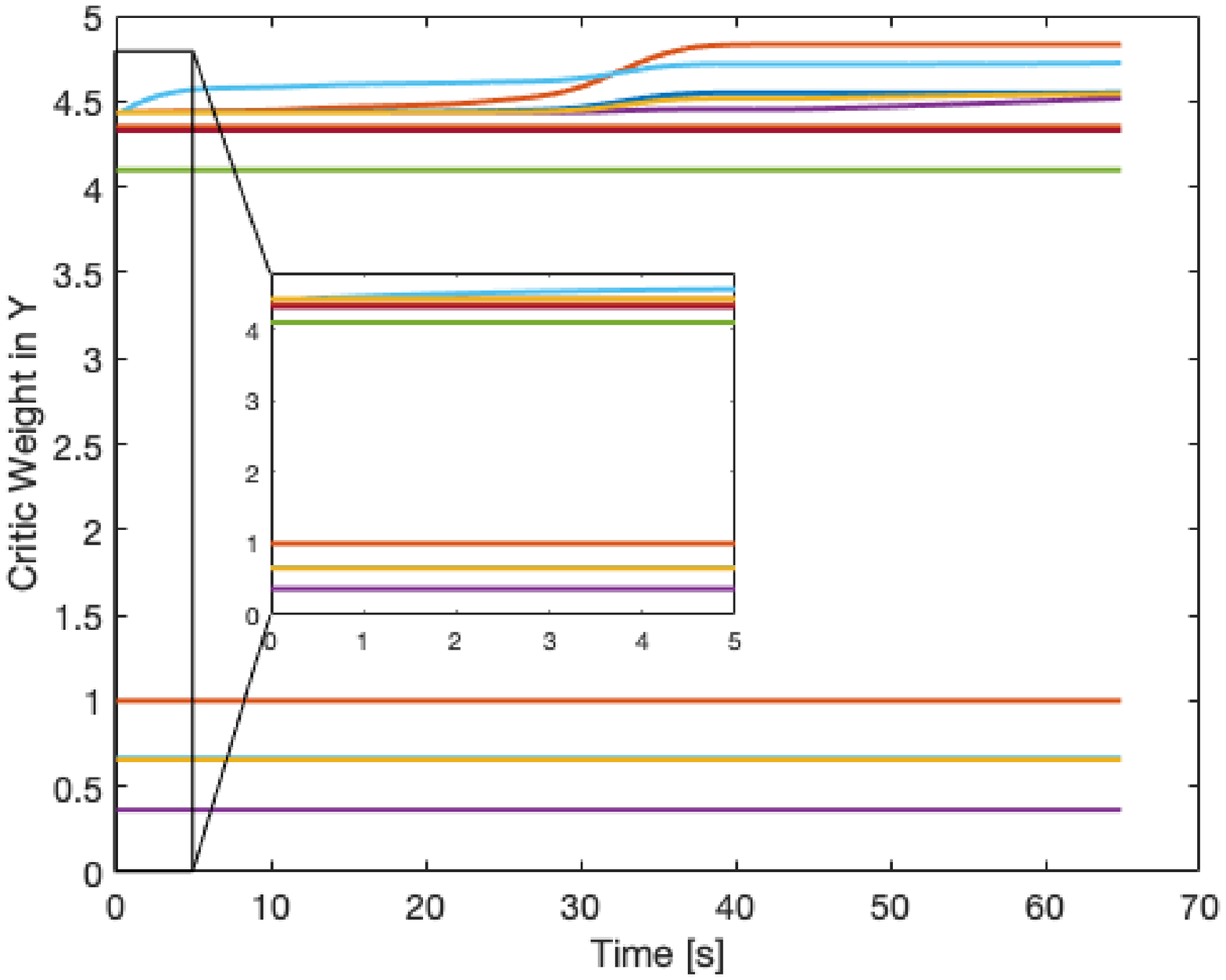}%
  }%
  \caption{Tracking critic weights of the proposed PI technique and the VI approach of~\cite{ICRA21} along the $x$ and $y$-directions\label{fig:tracking-critic-weights}} 
\end{figure*}

Another interesting observation about the proposed RLS-based PI algorithm is its faster critic weight convergence along the $x$ and $y$-directions when compared to its VI counterpart presented in~\cite{ICRA21}. \Cref{fig:tracking-critic-weights} shows that the weights of the former method converge in less than \SI{1}{\s} and are not much affected later by the dynamic disturbances throughout the simulation, unlike the VI algorithm. This reflects a higher ability to adapt to time-varying graph topology in the flock decision process, which is encoded in the PI method, rather than relying on a fully connected-graph topology as it is the case with the VI approach.

\section{Conclusion}
\label{conc}

A novel guidance mechanism based on policy iteration is introduced to solve a flocking motion problem in real-time without knowing the dynamics of the agents or those of the formation. This solution is complemented with an extended fuzzy system to ensure that the agents will stay close to each other and avoid colliding. Additionally, they interact simultaneously considering a time-varying graph topology in order to reach consensus on a common flock velocity.     	
The proposed approach employ model-free strategies and it continuously evaluates and updates the guidance policies without waiting for data batches unlike least squares and batch least squares regression methods.     
The policy iteration solution based on recursive least squares exhibited better convergence characteristics when compared with another reinforcement learning process that is based on value iteration.
A real-world robotics simulation software engine is employed to show the usefulness of the developed solution for a flock of Pioneer-3DX mobile robots.



\IEEEtriggeratref{8} 
\IEEEtriggercmd{\enlargethispage{-1.1in}} 
\bibliographystyle{IEEEtran}
\bibliography{Bib/ref.bib}

\begin{thebibliography}{10}
\providecommand{\url}[1]{#1}
\csname url@samestyle\endcsname
\providecommand{\newblock}{\relax}
\providecommand{\bibinfo}[2]{#2}
\providecommand{\BIBentrySTDinterwordspacing}{\spaceskip=0pt\relax}
\providecommand{\BIBentryALTinterwordstretchfactor}{4}
\providecommand{\BIBentryALTinterwordspacing}{\spaceskip=\fontdimen2\font plus
\BIBentryALTinterwordstretchfactor\fontdimen3\font minus
  \fontdimen4\font\relax}
\providecommand{\BIBforeignlanguage}[2]{{%
\expandafter\ifx\csname l@#1\endcsname\relax
\typeout{** WARNING: IEEEtran.bst: No hyphenation pattern has been}%
\typeout{** loaded for the language `#1'. Using the pattern for}%
\typeout{** the default language instead.}%
\else
\language=\csname l@#1\endcsname
\fi
#2}}
\providecommand{\BIBdecl}{\relax}
\BIBdecl

\bibitem{Grundel2007}
D.~Grundel, R.~Murphey, P.~Pardalos, and O.~Prokopyev, \emph{Cooperative
  Systems: Control and Optimization}.\hskip 1em plus 0.5em minus 0.4em\relax
  Springer Science \& Business Media, Jan. 2007, vol. 588.

\bibitem{Zhihua2008}
Z.~Qu, J.~Wang, and R.~A. Hull, ``Cooperative control of dynamical systems with
  application to autonomous vehicles,'' \emph{IEEE Transactions on Automatic
  Control}, vol.~53, no.~4, pp. 894--911, 2008.

\bibitem{Wurman2008}
P.~Wurman, R.~D'Andrea, and M.~Mountz, ``Coordinating hundreds of cooperative,
  autonomous vehicles in warehouses.'' \emph{AI Magazine}, vol.~29, pp. 9--20,
  03 2008.

\bibitem{ICRA1}
U.~Halder and B.~Dey, ``Biomimetic algorithms for coordinated motion: Theory
  and implementation,'' in \emph{2015 IEEE International Conference on Robotics
  and Automation (ICRA)}, 2015, pp. 5426--5432.

\bibitem{ICRA2}
J.~M. Soares, A.~P. Aguiar, A.~M. Pascoal, and A.~Martinoli, ``A distributed
  formation-based odor source localization algorithm - design, implementation,
  and wind tunnel evaluation,'' in \emph{2015 IEEE International Conference on
  Robotics and Automation (ICRA)}, 2015, pp. 1830--1836.

\bibitem{Speck2018}
C.~Speck and D.~J. Bucci, ``Distributed {UAV} swarm formation control via
  object-focused, multi-objective sarsa,'' in \emph{2018 Annual American
  Control Conference}, 2018, pp. 6596--6601.

\bibitem{gu2008using}
D.~Gu and H.~Hu, ``Using fuzzy logic to design separation function in flocking
  algorithms,'' \emph{IEEE Transactions on Fuzzy Systems}, vol.~16, no.~4, pp.
  826--838, 2008.

\bibitem{abouheaf2017flocking}
M.~Abouheaf and W.~Gueaieb, ``Flocking motion control for a system of
  nonholonomic vehicles,'' in \emph{2017 IEEE International Symposium on
  Robotics and Intelligent Sensors (IRIS)}.\hskip 1em plus 0.5em minus
  0.4em\relax IEEE, 2017, pp. 32--37.

\bibitem{reynolds1987flocks}
C.~W. Reynolds, ``Flocks, herds and schools: A distributed behavioral model,''
  in \emph{Proceedings of the 14th annual conference on Computer graphics and
  interactive techniques}, 1987, pp. 25--34.

\bibitem{herd}
A.~A. Paranjape, S.-J. Chung, K.~Kim, and D.~H. Shim, ``Robotic herding of a
  flock of birds using an unmanned aerial vehicle,'' \emph{IEEE Transactions on
  Robotics}, vol.~34, no.~4, pp. 901--915, 2018.

\bibitem{hu2010distributed}
J.~Hu and G.~Feng, ``Distributed tracking control of leader--follower
  multi-agent systems under noisy measurement,'' \emph{Automatica}, vol.~46,
  no.~8, pp. 1382--1387, 2010.

\bibitem{jadaliha2010adaptive}
M.~Jadaliha, J.~Lee, and J.~Choi, ``Adaptive control of multi-agent systems for
  finding peaks of unknown fields,'' in \emph{Dynamic Systems and Control
  Conference}, vol. 44182, 2010, pp. 623--630.

\bibitem{aerial}
S.~Oweis, S.~Ganesan, and K.~C. Cheok, ``Server based control flocking for
  aerial-systems,'' in \emph{IEEE International Conference on
  Electro/Information Technology}, 2014, pp. 314--319.

\bibitem{fish}
Y.~Jia and L.~Wang, ``Leader–follower flocking of multiple robotic fish,''
  \emph{IEEE/ASME Transactions on Mechatronics}, vol.~20, no.~3, pp.
  1372--1383, 2015.

\bibitem{AbouheafAuto14}
\BIBentryALTinterwordspacing
M.~I. Abouheaf, F.~L. Lewis, K.~G. Vamvoudakis, S.~Haesaert, and R.~Babuska,
  ``Multi-agent discrete-time graphical games and reinforcement learning
  solutions,'' \emph{Automatica}, vol.~50, no.~12, pp. 3038--3053, 2014.
  [Online]. Available:
  \url{https://www.sciencedirect.com/science/article/pii/S0005109814004282}
\BIBentrySTDinterwordspacing

\bibitem{AbouheafCDC13}
M.~I. Abouheaf and F.~L. Lewis, ``Multi-agent differential graphical games:
  Nash online adaptive learning solutions,'' in \emph{52nd IEEE Conference on
  Decision and Control}, 2013, pp. 5803--5809.

\bibitem{AbouheafIJCNN13}
------, ``Approximate dynamic programming solutions of multi-agent graphical
  games using actor-critic network structures,'' in \emph{The 2013
  International Joint Conference on Neural Networks (IJCNN)}, 2013, pp. 1--8.

\bibitem{sut92}
R.~S. Sutton, A.~G. Barto, and R.~J. Williams, ``Reinforcement learning is
  direct adaptive optimal control,'' \emph{IEEE Control Systems Magazine},
  vol.~12, no.~2, pp. 19--22, 1992.

\bibitem{AbouheafIRL2019}
\BIBentryALTinterwordspacing
M.~Abouheaf, W.~Gueaieb, and A.~Sharaf, ``Load frequency regulation for
  multi-area power system using integral reinforcement learning,'' \emph{IET
  Generation, Transmission \& Distribution}, vol.~13, no.~19, pp. 4311--4323,
  2019. [Online]. Available:
  \url{https://ietresearch.onlinelibrary.wiley.com/doi/abs/10.1049/iet-gtd.2019.0218}
\BIBentrySTDinterwordspacing

\bibitem{Sutton_1998}
R.~S. Sutton and A.~G. Barto, \emph{Reinforcement Learning: An Introduction},
  2nd~ed., ser. Second.\hskip 1em plus 0.5em minus 0.4em\relax Massachusetts:
  MIT Press, 1998.

\bibitem{Bertsekas1996}
D.~Bertsekas and J.~Tsitsiklis, \emph{Neuro-Dynamic Programming}, 1st~ed.\hskip
  1em plus 0.5em minus 0.4em\relax Massachusetts: Athena Scientific, 1996.

\bibitem{AbouheafCTT2015}
M.~Abouheaf, F.~Lewis, M.~Mahmoud, and D.~Mikulski, ``Discrete-time dynamic
  graphical games: Model-free reinforcement learning solution,'' \emph{Control
  Theory and Technology}, vol.~13, no.~1, pp. 55--69, 2015.

\bibitem{Abouheapolicy2017}
M.~Abouheaf and M.~Mahmoud, ``Policy iteration and coupled riccati solutions
  for dynamic graphical games,'' \emph{International Journal of Digital Signals
  and Smart Systems}, vol.~1, no.~2, pp. 143--162, 2017.

\bibitem{Singh2013}
H.~Singh, M.~M. Gupta, T.~Meitzler, Z.-G. Hou, K.~K. Garg, A.~M.~G. Solo, and
  L.~A. Zadeh, ``Real-life applications of fuzzy logic,'' \emph{Advances in
  Fuzzy Systems}, vol. 2013, pp. 1--3, 2013.

\bibitem{innocenti2007multi}
B.~Innocenti, B.~L{\'o}pez, and J.~Salvi, ``A multi-agent architecture with
  cooperative fuzzy control for a mobile robot,'' \emph{Robotics and Autonomous
  Systems}, vol.~55, no.~12, pp. 881--891, 2007.

\bibitem{6762934}
H.~Zhang, J.~Zhang, G.-H. Yang, and Y.~Luo, ``Leader-based optimal coordination
  control for the consensus problem of multiagent differential games via fuzzy
  adaptive dynamic programming,'' \emph{IEEE Transactions on Fuzzy Systems},
  vol.~23, pp. 152--163, Jan. 2014.

\bibitem{ICRA21}
S.~Qu, M.~Abouheaf, W.~Gueaieb, and D.~Spinello, ``An adaptive fuzzy
  reinforcement learning cooperative approach for the autonomous control of
  flock systems,'' in \emph{2021 International Conference on Robotics and
  Automation (ICRA)}, 2021.

\bibitem{Busoniu2010}
L.~Buşoniu, D.~Ernst, B.~De~Schutter, and R.~Babuška, ``Online least-squares
  policy iteration for reinforcement learning control,'' in \emph{Proceedings
  of the 2010 American Control Conference}, 2010, pp. 486--491.

\bibitem{Srivastava2019}
R.~Srivastava, R.~Lima, K.~Das, and A.~Maity, ``Least square policy iteration
  for ibvs based dynamic target tracking,'' in \emph{2019 International
  Conference on Unmanned Aircraft Systems (ICUAS)}, 2019, pp. 1089--1098.

\bibitem{Lewis2012}
F.~L. Lewis, D.~Vrabie, and V.~L. Syrmos, \emph{Optimal Control}.\hskip 1em
  plus 0.5em minus 0.4em\relax John Wiley \& Sons, 2012.

\bibitem{Yin2015}
B.~Yin, M.~Dridi, and A.~El~Moudni, ``Approximate dynamic programming with
  recursive least-squares temporal difference learning for adaptive traffic
  signal control,'' in \emph{IEEE Conference on Decision and Control}, 2015,
  pp. 3463--3468.

\bibitem{1315946}
Y.~Engel, S.~Mannor, and R.~Meir, ``The kernel recursive least-squares
  algorithm,'' \emph{IEEE Transactions on Signal Processing}, vol.~52, no.~8,
  pp. 2275--2285, 2004.

\bibitem{1163587}
D.~Lee, M.~Morf, and B.~Friedlander, ``Recursive least squares ladder
  estimation algorithms,'' \emph{IEEE Transactions on Acoustics, Speech, and
  Signal Processing}, vol.~29, no.~3, pp. 627--641, 1981.

\bibitem{1186866}
E.~Wilson, C.~Lages, and R.~W. Mah, ``On-line gyro-based, mass-property
  identification for thruster-controlled spacecraft using recursive least
  squares,'' in \emph{The Midwest Symposium on Circuits and Systems
  (MWSCAS-2002)}, vol.~2, 2002, pp. 1--4.

\bibitem{6529766}
S.~Shafigh, T.~Zia, and N.~Mouzehkesh, ``Wireless accelerometer sensor data
  filtering using recursive least squares adaptive filter,'' in \emph{2013 IEEE
  Eighth International Conference on Intelligent Sensors, Sensor Networks and
  Information Processing}, 2013, pp. 66--70.

\bibitem{5451430}
W.~Xu and F.~Liu, ``Recursive algorithm of generalized least squares
  estimator,'' in \emph{2010 The 2nd International Conference on Computer and
  Automation Engineering (ICCAE)}, vol.~3, 2010, pp. 487--490.

\bibitem{8641452}
A.~Rastegarnia, ``Reduced-communication diffusion {RLS} for distributed
  estimation over multi-agent networks,'' \emph{IEEE Transactions on Circuits
  and Systems II: Express Briefs}, vol.~67, no.~1, pp. 177--181, 2020.

\bibitem{YUSOF20111717}
\BIBentryALTinterwordspacing
R.~Yusof, R.~Z. {Abdul Rahman}, M.~Khalid, and M.~F. Ibrahim, ``Optimization of
  fuzzy model using genetic algorithm for process control application,''
  \emph{Journal of the Franklin Institute}, vol. 348, no.~7, pp. 1717--1737,
  2011, special issue on Modeling, Simulation and Applied Optimization.
  [Online]. Available:
  \url{https://www.sciencedirect.com/science/article/pii/S0016003210002310}
\BIBentrySTDinterwordspacing

\bibitem{7801936}
J.-W. Yeh and S.-F. Su, ``Efficient approach for {RLS} type learning in {TSK}
  neural fuzzy systems,'' \emph{IEEE Transactions on Cybernetics}, vol.~47,
  no.~9, pp. 2343--2352, 2017.

\bibitem{6083937}
J.-W. Yeh, S.-F. Su, and I.~Rudas, ``Analysis of using rls in neural fuzzy
  systems,'' in \emph{2011 IEEE International Conference on Systems, Man, and
  Cybernetics}, 2011, pp. 1831--1836.

\bibitem{hu2005adaptive}
C.-C. Hu, H.-Y. Lin, and J.-H. Wen, ``An adaptive fuzzy-logic variable
  forgetting factor rls algorithm,'' in \emph{2005 IEEE 62nd Vehicular
  Technology Conference, 2005}, vol.~3.\hskip 1em plus 0.5em minus 0.4em\relax
  IEEE, 2005, pp. 1412--1416.

\bibitem{1593689}
Y.~Xu, K.-W. Wong, and C.-S. Leung, ``Generalized rls approach to the training
  of neural networks,'' \emph{IEEE Transactions on Neural Networks}, vol.~17,
  no.~1, pp. 19--34, 2006.

\bibitem{aastrom2013adaptive}
K.~J. {\AA}str{\"o}m and B.~Wittenmark, \emph{Adaptive Control}.\hskip 1em plus
  0.5em minus 0.4em\relax Courier Corporation, 2013.

\bibitem{casteigts2010}
A.~Casteigts, P.~Flocchini, W.~Quattrociocchi, and N.~Santoro, ``Time-varying
  graphs and dynamic networks,'' in \emph{Ad-hoc, Mobile, and Wireless
  Networks}, H.~Frey, X.~Li, and S.~Ruehrup, Eds.\hskip 1em plus 0.5em minus
  0.4em\relax Berlin, Heidelberg: Springer Berlin Heidelberg, 2011, pp.
  346--359.

\bibitem{Olfati2006}
R.~Olfati-Saber, ``Flocking for multi-agent dynamic systems: Algorithms and
  theory,'' \emph{IEEE Transactions on Automatic Control}, vol.~51, no.~3, pp.
  401--420, 2006.

\end{thebibliography}

\end{document}